# A Proposal for a Cryogenic Experiment to Measure the Neutron Electric Dipole Moment (nEDM)

S.N.Balashov, K. Green, M.G.D. van der Grinten,
P.G.Harris, H.Kraus, J.M.Pendlebury, D.B.Shiers,
M.A.H.Tucker, D.L.Wark

For the
Sussex/RAL/Oxford/Kure/ILL
nEDM Collaboration

*This document is a copy of the original 2003 proposal for the construction grant for the CryoEDM Experiment at ILL, Grenoble. It is here made publicly available as a technical reference source for interested parties. It does not necessarily represent the final configuration of the experiment. Items pertaining to costs, personnel etc. have been removed.*

# Executive Summary

After almost 40 years of investigation, the problem of CP violation and its relation to the excess of matter over anti-matter in the observed Universe remains one of the core problems in particle physics and astrophysics. The recent observations of CP violation in the decay of B mesons (only the second observation of CP violation in the fundamental laws of physics ever seen) seem to confirm our picture of CP violation in the Standard Model of particle physics, but this only deepens the overall mystery as that CP violation is too small to explain the observed dominance of matter in the Universe. We must therefore search for CP violation arising from physics beyond the Standard Model. A particularly sensitive probe for such CP violation is offered by searches for (and hopefully soon measurements of) particle electric dipole moments, and in particular (for the purposes of this proposal) on the static electric dipole moment of the neutron (nEDM). Current world-leading limits on the nEDM arising from the existing Sussex/RAL/ILL experiment are already one of the tightest available constraints on models of physics beyond the Standard Model, in particular they place strong constraints on the allowable form of supersymmetric models which are for many the theoretical favourite for the next great theory of physics, and which form the scientific justification for such vastly larger projects as the LHC at CERN.

The existing nEDM experiment has about reached its limiting sensitivity, largely because the source of the ultra-cold neutrons (UCN) at the experiment's home at the ILL laboratory in Grenoble, France cannot provide enough neutrons to reach the experiment's limiting systematic sensitivity. We therefore propose to build an entirely new experiment, based on a new method for the production of UCN by the downscattering of cold neutrons in superfluid liquid helium, which we have recently quantitatively demonstrated for the first time. An "atomic clock" using these UCN as the active element would then be built within the LHe. Frequency shifts of this clock as a function of a strong applied electric field are then a signature of a non-zero nEDM. The properties of LHe would allow all the physical parameters which determine the measurement sensitivity of such an experiment to be improved relative to the existing experiment, in most cases by very significant factors, while at the same time allowing the systematic uncertainties in the experiment to be controlled. The proposed experiment is designed to take advantage of this combination to produce an experiment which would be sensitive to a nEDM at the level of $1.7 \times 10^{-28}$ $e$ cm, which is an improvement of two orders of magnitude in sensitivity compared to our current experiment and covers the region in which the answer is predicted to lie by natural supersymmetric models.

The exact running time necessary to reach this level of sensitivity with currently available cold neutron facilities is difficult to estimate prior to the measurement of the precise performance of the apparatus proposed here, and could be as long as 5 years. This proposal therefore concentrates on the 3-year capital construction phase of the project when an experiment of this sensitivity would be built. As part of the construction and commissioning of this experiment an nEDM experiment with a sensitivity of $10^{-27}$ $e$ cm would be performed (a full order of magnitude more sensitive than the existing limit). Late in the period of this grant, when enough experience had been gained of the performance of the equipment, we would return to PPARC for a separate grant for the exploitation phase and any proposed further development.

# Table of Contents

# 1. Introduction

## 1.1 CP Violation and the neutron EDM

The detailed experimental study of CP violation in fundamental physics dates from the 1964 paper reporting the first observations of CP violation in the decay of neutral K mesons (J.H. Christenson et al., PRL 13, 138-140, 1964). This observation gained even greater importance when the deep connection between CP violation in the fundamental laws of physics and the observed matter-anti-matter asymmetry of the Universe was first pointed out in 1967 in a famous paper by Andre Sakharov (A.D. Sakharov, JETP Lett. 5, 24-27, 1967). Recently measurements of CP-violation in the decay of B mesons by the BaBar and Belle collaborations have confirmed the Standard Model explanation of the CP violation seen in the neutral K system in terms of a CP-violating phase in the CKM matrix. This is in one sense satisfying, as it demonstrates the power of the Standard Model to explain a wide range of phenomena, however it is also frustrating as we know that the CP violation in the Standard Model is too small (by many orders of magnitude) to explain the observed baryon asymmetry in the Universe. We are therefore left to find another source of CP violation in fundamental physics if we are to understand as basic a question as "Why does the matter in the Universe exist?".

The ideal place to look for CP violation from physics beyond the Standard Model is to look in a place where the Standard Model contribution is small, but where contributions from non-SM processes tend to be large enough to be measured. The static electric dipole moment of the neutron (nEDM) is just such a quantity. Neutrons are neutral bodies, but it is possible for their positive and negative internal charge distributions to be slightly displaced, which would result in a non-zero nEDM. Such a nEDM would violate CP (this can be seen intuitively in that it obviously violates T, as under T the charge distributions are invariant and therefore the nEDM would not change sign, however the spin reverses so the relative orientation of the nEDM and the only "direction" in the problem, the direction of the neutron spin, would change: thus it violates T and under CPT it must violate CP as well). A similar argument shows that it violates parity as well, so we are looking for a T-odd, P-odd observable.

The Standard Model contribution to the nEDM is very small, in fact so small ($\sim 10^{-31\text{-}32}$ $e$ cm) that it would be extremely difficult to measure. However most extensions to the SM produce much larger predictions for the nEDM, and in particular, a nEDM arises at the one-loop level in supersymmetric models leading to a natural scale for the nEDM in such models at $\sim 10^{-23}$ $e$ cm. As will be described below, this has already been ruled out by our existing experiment, a situation which is described in the literature as the "SUSY CP problem". There is an extensive literature on this subject (our best current limit was published in 1999 and has already been cited 140 times as counted by SPIRES), and no room here to give a thorough review. A number of models have been proposed to manoeuvre around the experimental constraints – perhaps for some reason all the CP violating supersymmetric phases are small (it is worth noting that the QCD Lagrangian would most naturally produce an nEDM of $\sim 10^{-16}$ $e$ cm, so some mechanism has very highly suppressed that phase), perhaps the masses of the sparticles are very large, perhaps there are cancellations or fine tuning. Unless there is some symmetry principle that we are missing all of these explanations look rather unnatural, and all would be severely strained by a further improvement in the sensitivity of our measurement. It is worth noting here that the same applies to the EDM of the electron, where both the experimental limit and the theoretical prediction lie 1-2 orders of magnitude below the nEDM value, leading to similar but complementary constraints on the underlying physics. It is repeatedly stressed in the literature that if supersymmetry is a property of nature then particle EDMs should lie not far below the current experimental limits, and that measurements of particle EDMs provide the most stringent bounds on current model building. Of course, perhaps supersymmetry is not a property of nature, and in that case it is worth pointing out that particle EDMs arise in most theories of physics beyond the SM (except for theories that have been specifically constructed to tune them away) – for examples left-right symmetric models and multiple Higgs models also predict values near the current experimental limits. The sheer number of papers trying to find ways to argue around nEDM constraints on new physics shows how valuable EDM measurements are at providing insight into the structure of physics beyond the Standard Model, and how valuable a further improvement in the sensitivity of our experiments would be at constraining such models. Of course even more important would be if one of those models were right, in which case we can expect to make a discovery if we are able to push the sensitivity of our experiment by another 1-2 orders of magnitude. We shall argue below that given the resources requested in this proposal we would do exactly that, and on a time scale where the results would be available well in advance of first results from the LHC.

The science case for this experiment, especially given its very low cost, seems compelling and has lead to its inclusion in Science Committee's recommended science programme and to favourable comment from all the advisory panels which have reviewed it. One might well then ask: Why nobody else has decided to do this experiment? The answer is, firstly, that even though it is not expensive it is very complex and technically challenging and the UK experience in this area is a very considerable asset. The other answer, however, is that others have decided to do it. A major programme has begun at PSI with involvement of the PNPI group who made the last non-Sussex/RAL measurement to produce a UCN source based on solid D2 (which is more appropriate for a spallation neutron source like PSI) leading to an nEDM experiment of similar claimed sensitivity to ours. There is also a large American group proposing an experiment rather similar to ours to be based at the new SNS facility at Oak Ridge. This group's proposal has 36 names on it, asks for $11M, and proposes to make measurements starting in 2008 with eventual sensitivity pushing down into the ~$10^{-29}$ $e$ cm region. There is also a group at the new research reactor in Munich who are planning on proposing a nEDM experiment when (if?) that reactor is ever started. We therefore have severe competition over whom we have a lead, but not a lead we will keep without proper support.

## 1.2 Measuring the nEDM

The basic idea used for the measurement of the nEDM both in our existing room-temperature experiment and in our proposed cryogenic experiment is based on the Ramsey oscillator invented by Norman Ramsey (who, in addition to winning the Nobel Prize for this invention, was a former collaborator of ours who made pioneering nEDM measurements, see section 1.4). The technique is very similar to that used in an atomic clock, in fact our experiments are basically "atomic clocks" using neutrons instead of atoms, with the signature of a non-zero nEDM being a shift in the frequency of the neutron clock under an applied electric field. Polarised neutrons are first prepared with their spin in, say, the +z direction. A short (~2s) oscillatory magnetic field is then used to precess their spins by 90° onto, say, the x axis. A constant magnetic field $B_o$ is then applied in the z direction which makes the spin precess about the z axis at the Larmor frequency . After a much longer time interval T a second pulse of the oscillatory field is again applied. If an exact integral multiple of the Larmor period has elapsed before this second pulse the spin will again be along the x axis, and the pulse will precess the spin down to the –z axis. A subsequent measurement of the neutron polarisation would then show them to be 100% polarised with the spins opposite to the original direction. If, however, T is ½ a Larmor period more than an integral multiple the spins will be along the –x axis and the short pulse will actually align them back with the +z axis, returning their polarisation to the initial state. This is shown in Figure N, which shows the Ramsey resonance curve – the number of neutrons as a function of T transmitted through a polariser which transmits the spin component oppositely aligned to the original spin. Measuring this curve gives sensitivity to shifts of a small fraction of the Larmor period.

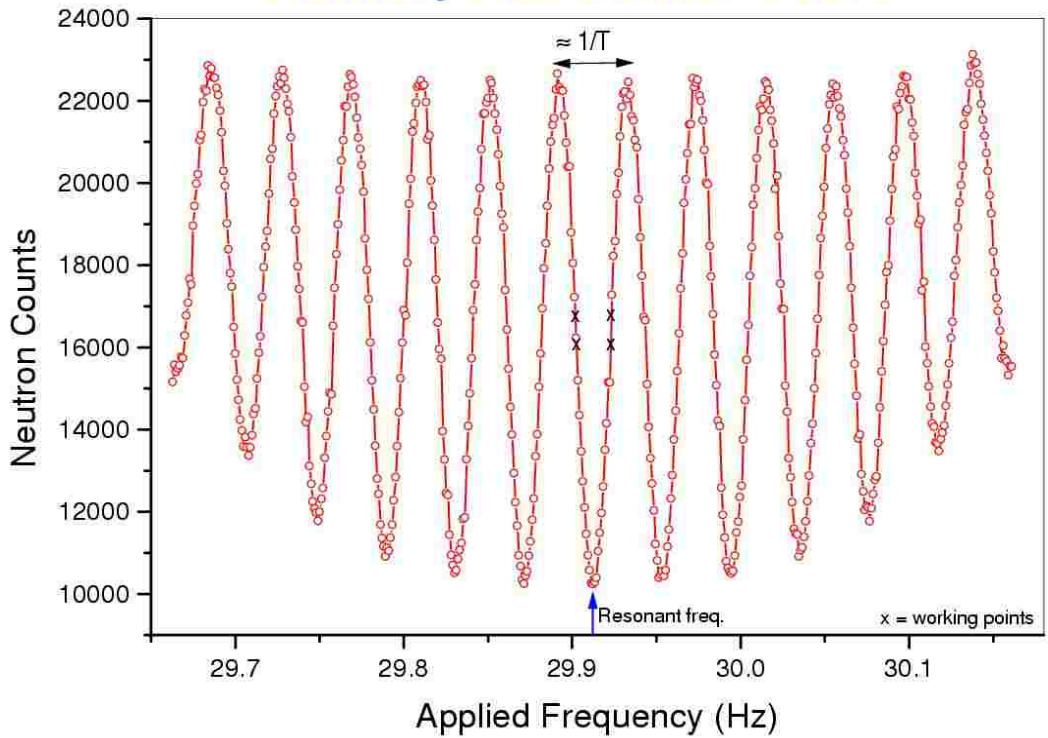

How can this be adapted to a measurement of the nEDM? The Larmor frequency for a neutron in a magnetic field $B_o$ is given by $2|\mu_n|B_o$ . If the neutron has an EDM $d_n$ an electric field $E$ applied parallel to $B_o$ will produce a shift in this frequency of $\pm 2d_n \cdot E$, where the sign depends on the relative signs of $E$ and $B_o$. If the sign of $E$ is flipped there will therefore be a shift of $4d_n \cdot E$ in the frequency, which can be measured by the above technique. In practice we are looking for the smallest measurable shifts in the frequency, so we tune the experiment to one of the half-height points near the centre of Figure N (where the slope is the greatest). The neutrons are prepared in a spin-polarised state, introduced into a volume with fields $E$ and $B_o$, and the Larmor frequency measured as above. The last step is to count the number of neutrons $N_1$ and $N_2$ that finish in the two spin states (up or down) relative to a holding magnetic field. The storage cell is operated on a batch cycle principle: (a) fill with polarised neutrons, (b) carry out the magnetic resonance, and (c) empty, spin analyse and detect to obtain $N_1$ and $N_2$. Then the relative direction of **E** and **B$_o$** are flipped, and the cycle repeated (actually many cycles are run with each field orientation, but conceptually that doesn't matter). Calling the numbers with fields parallel $N_{i\uparrow\uparrow}$, and the numbers with fields anti-parallel $N_{i\uparrow\downarrow}$, $d_n$ can be derived directly from:

$$1. \qquad d_n = \frac{(N_{1\uparrow\uparrow} - N_{2\uparrow\uparrow} - N_{1\uparrow\downarrow} + N_{2\uparrow\downarrow})\hbar}{2\alpha ETN}$$

where $N$ is the total number of neutrons counted and $\alpha$ is the neutron polarisation product (the product of the polarisation and the analysing efficiency, which determines the visibility of the central fringe). An important quantity is the error in this determination arising from counting statistics alone. This is given by:

$$2. \qquad \sigma(d_n) = \frac{\hbar}{2\alpha ET\sqrt{N}}$$

From this description the important elements in an nEDM experiment of this type follow:

i.) A source of neutrons. The sensitivity depends directly on the square root of the total number of neutrons detected.

ii.) A way to polarise the neutrons, and to analyse their polarisation at the end. The sensitivity depends on $\alpha$, the product of the polarisation and analysing efficiency.

iii.) A constant magnetic field **B$_o$**. Any inhomogeneity in this field will cause neutrons in different parts of the volume to precess with different frequencies, destroying the coherence and reducing $\alpha$ and reducing the sensitivity.

iv.) A way to control the magnetic environment to reduce and monitor stray environmental B fields. Either stray fields must be reduced to the point that they do not affect the measurement or they must be monitored and corrections applied during analysis.

v.) An electric field **E**. The sensitivity depends linearly on the magnitude of this field, which should therefore be as great as possible. However any leakage currents arising from the application of this field will produce B fields which can produce systematic uncertainties or even a false EDM signal. Control of these currents is therefore essential.

## 1.3 The production and bottling of UCN

Recent nEDM experiments rely on the rather unusual property (for a particle) of neutrons at very low energies (called ultra-cold neutrons, or UCN) that they can be literally bottled, i.e., placed in a container like the molecules in a gas. This arises because UCN have wavelengths long compared to the spacing between nuclei in a solid, and therefore they do not interact with individual nuclei but instead interact macroscopically with the medium. The resulting potential energy depends on the type of nuclei involved and to a lesser extent on the material's structure, but the effect is that for many materials the neutrons actually get an effective refractive index in the material that is less than one which leads to total external reflection at the interface. Below some critical energy of incidence (called the Fermi potential) this total external reflection occurs at all angles, and thus below that energy a neutron in a closed container is trapped. The time for which it is trapped is limited by a number of

processes – upscattering from the walls (where a phonon from the much walls, which are usually much hotter than the neutrons, causes the neutron to gain sufficient energy to escape the trap), capture on the nuclei of atoms on, or just within, the surface of the bottle, and eventually (if these other losses are small enough) the free decay of the neutron. For the purposes of our experiment another time is also important, which is the time it takes for the an initially polarised neutron to de-polarize (which can also occur via a number of processes and depends critically on the nature and magnetic properties of the bottle material, and on the homogeneity of $\mathbf{B_o}$) – we must try to make this time much longer than the bottling time in order to avoid losing sensitivity.

Currently we are using UCN from the TGV source at the ILL in Grenoble (this source is called PF2 at ILL). The source is a neutron turbine. Cold neutrons of ~50 m/s are produced with a liquid $D_2$ moderator near the core of the ILL reactor and then guided through a nearly vertical guide up to the turbine. These neutrons reflect several times off of the rapidly moving (retreating) nickel turbine blades, imparting momentum to the turbine and slowing down in the process to the point that the low energy tail can be bottled. This was, and still is, the most intense source of UCN in the world, currently producing a UCN number density of approximately 25 UCN $cm^{-3}$. For our next generation experiment we intend to no longer use this source, but to make UCN within our own apparatus with much higher densities (see section 2.1 below).

## 1.4 A brief history of nEDM measurements

Experiments to measure a neutron EDM have been carried out since 1950; many years before the discovery of CP violation in the $K^0$ decay. The figure shows the steady progress that has been achieved compared to the order-of-magnitudes of the predictions from various models. The techniques used in all the experiments have been conceptually similar to the above description – they all seek to measure a shift in a nuclear magnetic resonance (NMR) line of a free neutron when subjected to a strong electric field. The early experiments (Purcell and Ramsey 1950; Smith et al 1957; Miller et al 1967; Dress et al 1968 & 1977) used beams of neutrons when the neutrons were under observation for a few msec. An intrinsic problem with all these experiments was the so-called **Exv** effect when the motion of the neutron induces a magnetic field in it's rest frame and hence a precession of spin which is interpreted as a false nEDM signal. This effect is so severe that the most recent beam experiment ( *Dress et al Phys. Rev. D159 (1977)* ) reported $d_n < 3 \times 10^{-24}$ e cm which was an error 10x its statistical limit. More recently, techniques for storing UCN for many hundreds of seconds have been developed and the **Exv** problem has been essentially eliminated. As will be described in more detail below, the RAL/Sussex group has been in the forefront of these measurements (Pendlebury et al 1964; Smith et al 1990; Harris et al 1999) using facilities at the ILL in Grenoble. For a while similar experiments also took place at PNPI, Gatchina (Alterev et al 1986 & 1992 & 1996) but these ceased to operate in 1996. In order to give a feel for the current state of the art, the three most recent results are: $d_n = +(2.6 \pm 4.0 \pm 1.6)\times 10^{-26}$ *e* cm by PNPI in Russia [*Altarev et al; Atom. Phys. Nuclei 59 (1996) 11525*] and our group's two most recent measurements: $d_n = -(3 \pm 2 \pm 4)\times 10^{-26}$ *e* cm [*Smith et al; Phys. Lett. B234 (1990) 191*] and $d_n < 6.3 \times 10^{-26}$ *e* cm [*Harris et al; Phys. Rev. Lett. 82, (1999)904* ]. To give a feel for the sensitivity this most recent limit represents, imagine a neutron as made of two almost coincident spheres with charges +e and –e. Then an edm of $10^{-26}$ *e* cm corresponds to separating the centres of those spheres by $10^{-13}$ of the neutron's diameter, or, if you picture the neutron as being expanded to be the size of the Earth, separating the centres of the two charged spheres by about a micron. We will now give a brief account of how the current limits were obtained, and then go on to describe the proposed new experiment.

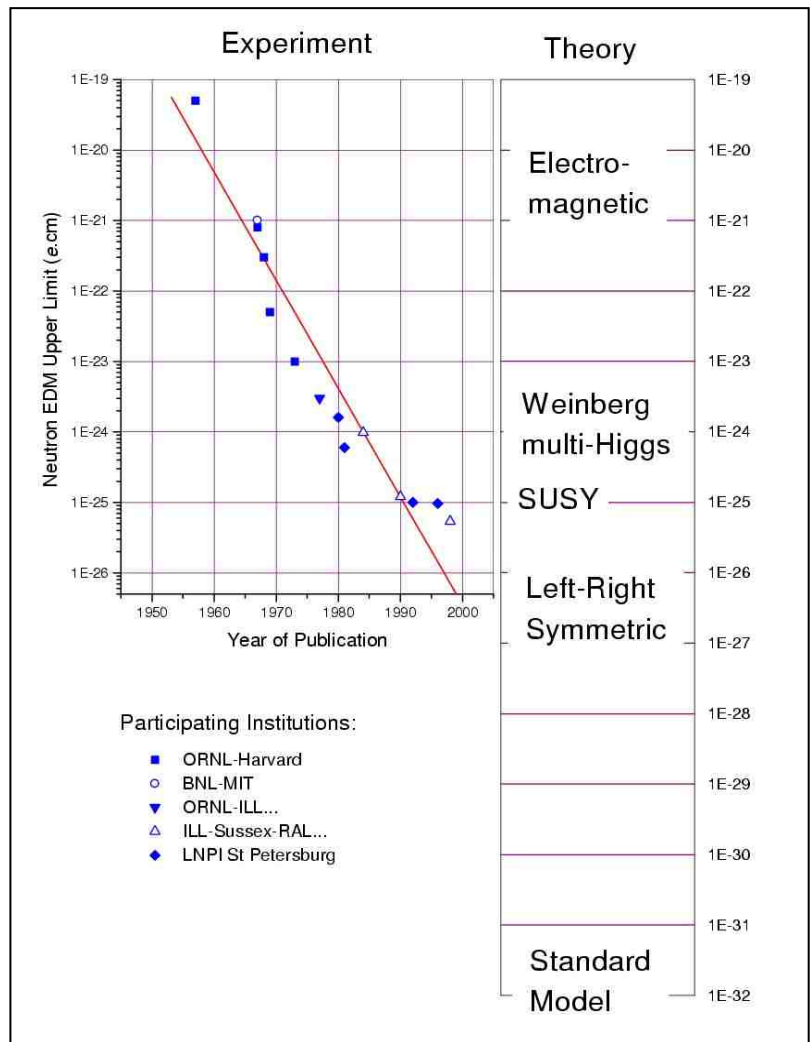

## 1.5 The Room Temperature Sussex/RAL/ILL nEDM Measurement.

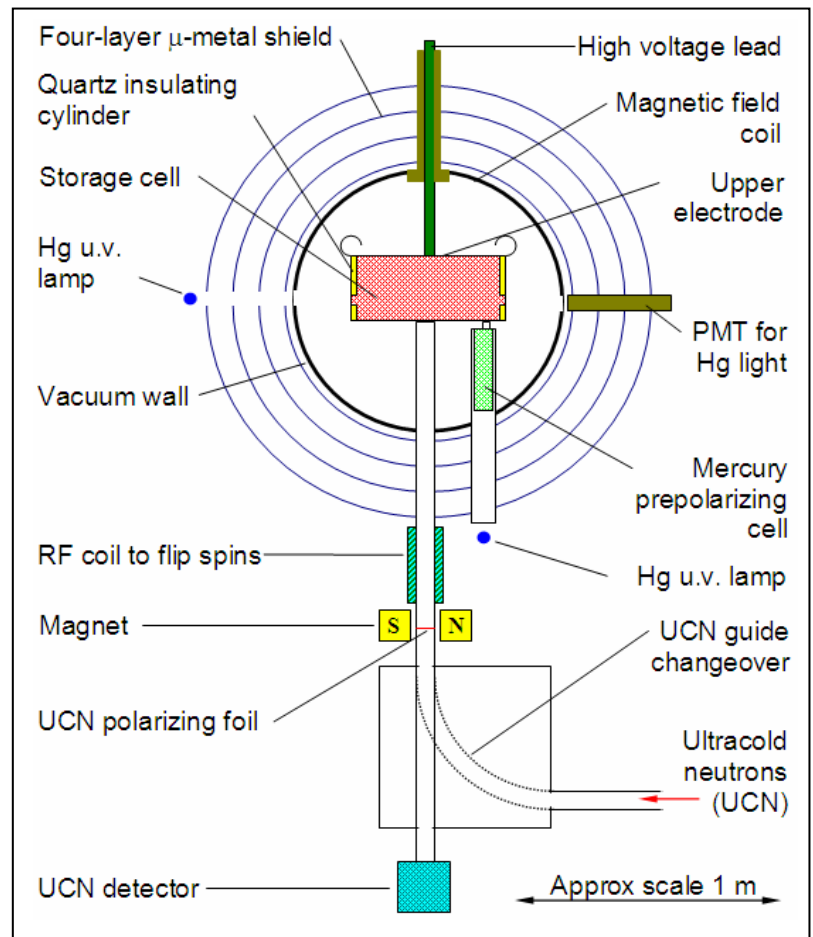

A diagram of the current experiment is shown in Figure K. UCN are admitted through a thin iron foil with an applied magnetic field. This reflects one spin component of the neutrons while transmitting the other, and thus polarised neutrons fill the storage cell (which is closed off with a door once filled). After the Ramsey cycle the door is opened and the neutrons in the proper polarisation state fall down to the UCN detector. After they are counted the RF coil is used to flip the spins of the other polarisation state (which have been trapped above the foil) and they are counted as well, doubling the statistics.

The apparatus has now been developed to the point where data is taken with $E = 12$ kV/cm; $T = 130$ sec; $\alpha = 0.64$ and $N = 14000$ neutrons per batch; with each batch cycle taking about 210 s. From one day of data, therefore (and allowing for pauses between runs), $\sigma(d_n)$ is about $2\times10^{-25}$ $e$ cm. The major advance which separates this nEDM experiment from all previous ones is its use of a cohabiting Hg magnetometer. A small amount ($3\times10^{10}$ atoms/cm$^3$) of $^{199}$Hg is polarised by optical pumping and then stored simultaneously in the same cell with the neutrons. The Larmor precession of the Hg can be monitored by observing with a photomultiplier (PMT) the sinusoidally varying transmission of circularly polarised light from a Hg lamp, and it can thus be used to monitor the magnetic field. Small temporal variations in the field arising from the magnetic environment can then be corrected at approximately the 2 nG level, increasing the sensitivity of the nEDM measurement by about a factor of 30. Data taken since 1999 have resulted in an improvement in sensitivity from the previously published value of $d_n=(1.9\pm5.4)\times10^{-26}$ $e$ cm cited above to a new value (preliminary and unpublished, and therefore confidential) of $d_n = -(0.7 \pm 1.5)\times10^{-26}$ $e$ cm, an improvement in sensitivity of more than a factor of 3.

The full systematic uncertainty analysis is not yet completed, and in particular there are systematics associated with known field gradients in the vicinity of the cell which will require a few further reactor cycles to fully characterize. This small amount of additional running (which will end in October 2003) and the final analysis of the data from this experiment are the subject of Work Package 1. Further details can be found in the appended report, in the papers cited therein, and in the Work Package 1 description in the Annex.

## 2. A Next Generation Experiment

How will we do better? Given that the existing experiment is not systematically limited we must improve one or more of the 4 quantities in the denominator on the right side of Equation 2 if we are going to improve the sensitivity – we need a higher electric field E, a higher polarisation product $\alpha$, a longer storage time T, more neutrons N. The experiment we have planned will actually improve all of these parameters. UCN will be produced by downscattering a cold neutron beam in 0.5 K liquid helium, and then be transported to a separate chamber where the EDM measurement will be made without the neutrons ever leaving the LHe. As will be described in section 2.1 this new production mechanism for UCN will produce 100x greater densities than currently available. LHe is actually a better insulator than vacuum, which should allow between 2 and 6 times higher values of E. The downscattering process should retain the nearly 100% polarisation which can be achieved in the cold neutron beam, thereby increasing the value of the polarisation product, while the ultra-cold and clean environment should allow storage times much longer than at present. The combination of all of these factors will improve the sensitivity of the experiment by an amount approaching two orders of magnitude, as will be discussed in more detail in the following sections.

## 2.1 A New Method for UCN production

A new method for the production of UCN with the potential to reach far higher densities than available from the ILL neutron turbine was first proposed in 1977 by

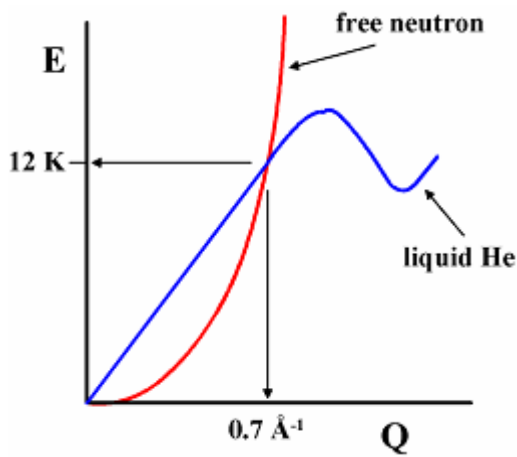

Bob Golub and one of us (JMP). This method relies on the properties of superfluid liquid helium (sLHe), specifically, on the dispersion curve as shown in Figure M. This plots the energy vs. the momentum for a free neutron (the red curve), which is of course just a parabola, and the energy vs. momentum for phonon excitations in the LHe (the blue curve). The properties of superfluid LHe are such that these two curves cross at a momentum corresponding to a UCN wavelength of 8.9Å. A neutron of that energy can therefore lose all of its energy and momentum by scattering off of a helium atom, essentially coming to a dead stop (and thus becoming a UCN). In order to suppress the inverse process the LHe must be cooled to very low temperatures, but at temperatures below 0.8 K high densities of UCN can be generated within the LHe.

Previous experiments have seen this process (Phys.Lett.**A125**,416(1987)), but our prototype experiments at the ILL with the apparatus discussed below have for the first time quantitatively verified that UCN production takes place at a rate consistent with theoretical prediction and is dominated by single-phonon scattering processes. This is shown in Figure Q, which plots the UCN density produced as a function of the wavelength of the incoming cold neutron beam. The observed production rate of $(0.91\pm 0.13)$ UCN/cm$^3$/sec from a beam of $2.6 \times 10^7$ n/cm$^2$/s/Å at 8.9 Å was very close to that expected for UCN production through a single phonon production process. These measurements give us confidence that using the existing cold neutron beam at the H53 position at ILL we will be able to produce UCN densities a factor of 100 higher than available for our current room-temperature experiment.

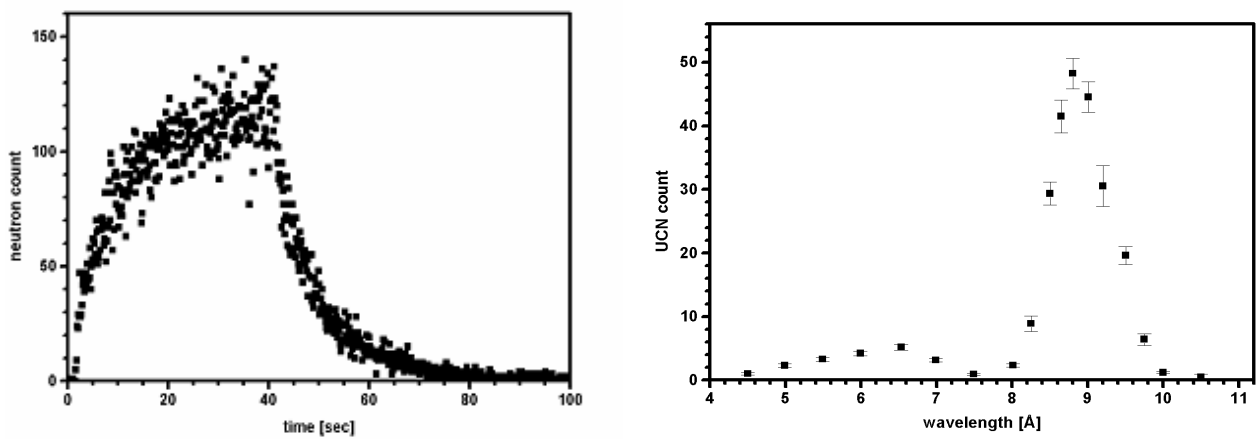

## 2.2 Overview of the Cryogenic nEDM Experiment

The experimental setup as originally envisaged is shown in Figure X (some planned layout changes will be described in the next section). The figure shows the 6 major sub-sections of the apparatus: Cooling Towers I and II, the UCN detection volume which is shown below Cooling Tower II on the diagram, the UCN production volume (which is represented on the diagram by the horizontal guide section connecting Cooling Towers I and II), the Ramsey Chamber, and the magnetic shields that surround it. Not shown are the enormous pumps which drive the actual cooling process. Cooling Towers I and II contain the bulk of the apparatus for reaching low temperatures. This is done in 4 stages. Firstly the apparatus is cooled by boil off of liquid $N_2$, and then by LHe, to reach 4 K. A He reservoir in Tower I is then pumped to further cool the He to 1.3 K. The final cooling is done by a closed-cycle $^3$He loop (also in Tower I) which cools the He to its final temperature of 0.4-0.7 K. A separate closed-cycle cooling system in Tower II is used to cool the magnetic shields so as to reduce the heat load on Tower I. The He within the UCN production, detection, and Ramsey chamber sections is actually isotopically purified $^4$He, which is produced by filling these sections through a superleak near the bottom of Tower I (which preferentially passes $^4$He and blocks $^3$He, which with its high neutron capture cross-section would shorten our neutron trapping time to a few seconds). The magnetic shields surrounding the Ramsey Chamber consist of three outer layers of mu-metal with an inner superconducting shield and a superconducting solenoid (the latter used to produce the $B_o$ field which is used in the Ramsey cycle). Superconductors are, of course, extremely effective shields against exterior magnetic field fluctuations, but the SC shield does not completely surround the chamber. This makes the monitoring of the fields which penetrate into this region a high priority (see section 2.2.4). In addition to these field monitors and the final design of the magnetic shielding of the UCN production and detection volumes and the Ramsey Chamber are the main components to be built for this proposal and will be described in more detail below. Cooling Towers I and II, the large pumps which drive them, the large vacuum chamber and shields, and most of the neutron guides and the cryostats around them and the UCN detection chamber are not asked for as part of this proposal because they already exist and are in use. They were supplied by our Japanese collaborator, Prof. Hajime Yoshiki, using a grant from the Japanese government. Cooling Tower II and the magnetic shields are currently in a laboratory at Sussex (refurbished using JIF funds), while Tower I and the UCN production and detection volumes are being used for tests at ILL.

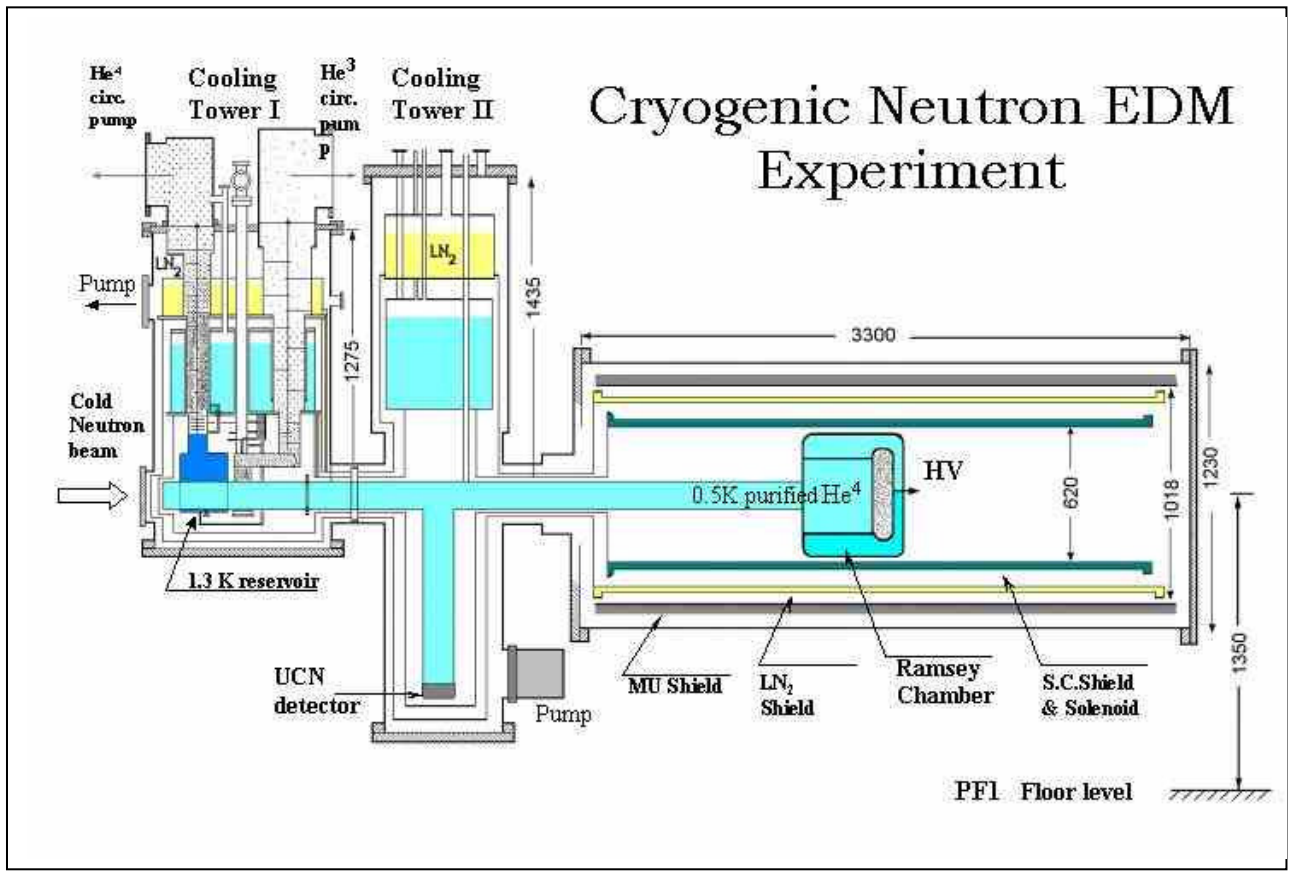

## 2.2.2 The cryostat system and neutron beamline

The first requirement for an experiment as described above is a source of 8.9Å neutrons (which is within the energy range called "cold" neutrons). As discussed in our earlier SoIs, the optimal solution in the long run would be to construct a new cold neutron beamline (with a higher flux of 8.9Å neutrons than currently available) at the ILL specifically dedicated to our experiment (this will be discussed below in section 2.3). However the PF1 position on the H53 beamline at ILL already provides a sufficient flux of cold neutrons to allow us to produce a cryogenic experiment with a substantially higher sensitivity than our existing experiment (see section 2.2.7 below), and in fact the flux potentially available at PF1 appears to be somewhat higher (perhaps a factor of 6) than we thought it would be when we wrote our SoIs. The collaboration therefore felt that before proposing to PPARC that a new beamline be constructed we should first demonstrate that a fully cryogenic experiment actually works, and determine experimentally whether the systematics in such an experiment are really low enough to make the additional statistics that would be available from a new beamline worth the money. This proposal is therefore structured around the PF1 position (although see sections 2.3 and 5).

The layout of the experimental elements actually proposed for the PF1 position is shown in Figure Z. As described in Work Packages 3 and 4 this will require new equipment and the modification of existing equipment. Firstly, the current H53 beamline ends at the shutter, and our cold neutrons therefore travel in free flight to our cryostat, with a corresponding reduction in the effective flux. We therefore need funds for high-quality neutron guides to link to our apparatus (see Work Package 3).

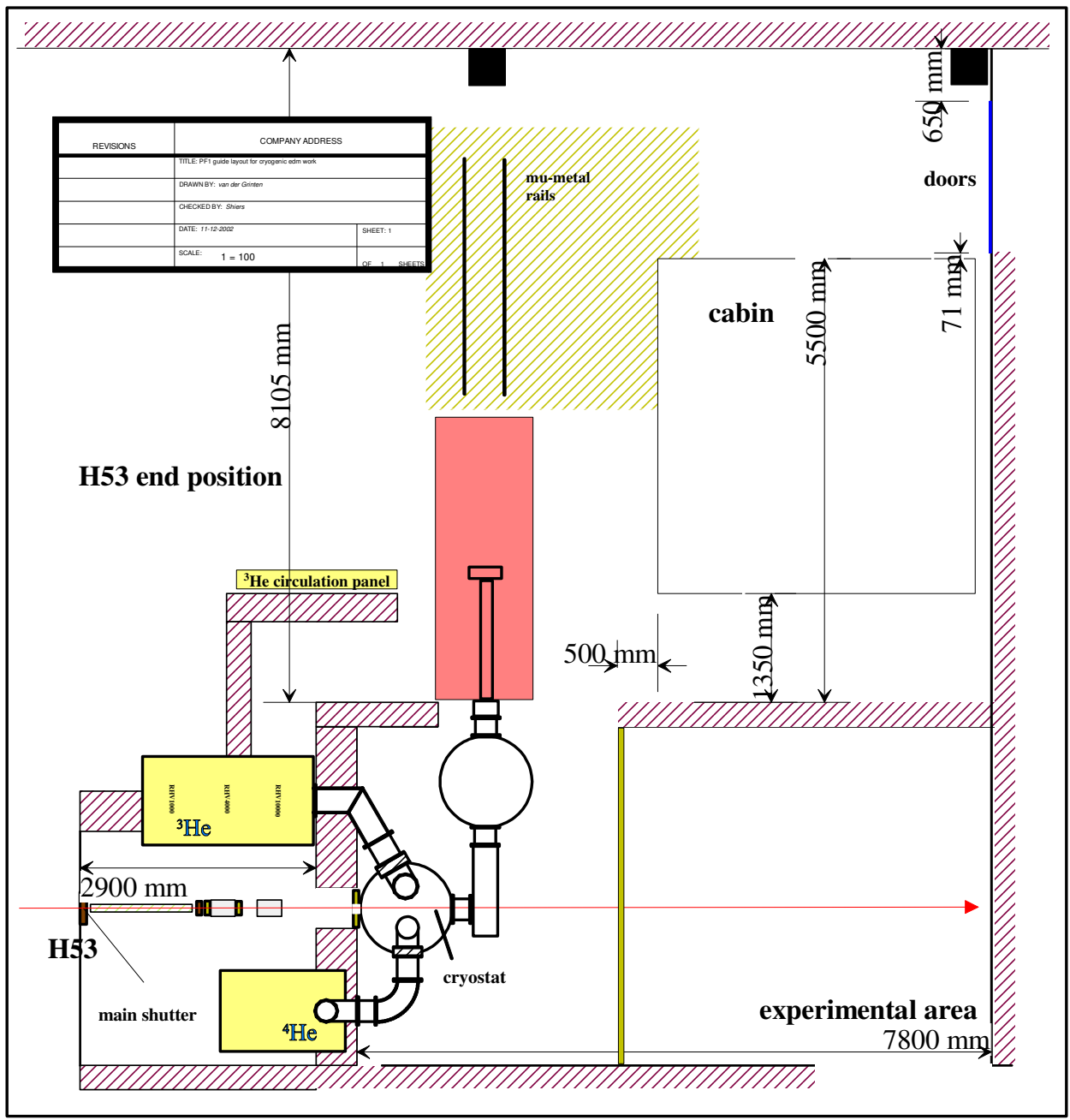

We also need a velocity selector and polariser (we are currently using borrowed equipment, but will not have permanent access to it).

## 2.2.3 UCN production volume and transfer

In Work Package 4 we describe changes to the layout of the cryostats and the UCN production volume in order to fit into the PF1 position and to maximize the stored neutron density. Figure Z shows the modification of the cryostats to an "L" shape, which removes the bulk of the equipment from the cold neutron beam and therefore from the problematic activation which it causes. It also allows us to operate the equipment in such a way as to simultaneously produce UCN in the production volume while making the measurement of the nEDM of the last batch of neutrons produced, thereby increasing our efficiency. In the picture the magnetic shields are the pink block near the middle of the picture, Tower II is the circle below the shields, and Tower I is below that (with the two large pumps shown as the yellow boxes). We request funds for the modifications necessary for the "L"-shaped arrangement (new guides to connect the production volume to the Ramsey Chamber), and also for extending the length of the UCN production region. An important design consideration in all of this is that we intend to polarise the incoming cold neutrons, which retain their polarisation upon downscattering (as verified for the first time in test experiments made at ILL, where preliminary results show that 93±10% of the initial polarisation is retained). Once polarised the neutrons must remain in a magnetic field to define the polarisation direction to prevent undue depolarisation. We therefore must have coils to produce this field built into the transfer guides and request funds for this purpose. Also needed are the valves which we open and close to move neutrons from the production volume into the Ramsey Chamber.

## 2.2.4 The Ramsey Cell

In Work Package 5 we outline the resources needed to produce the Ramsey Chamber for the new experiment. A schematic of the design is shown in Figure Y. The chamber is quite a bit more complicated than the one used in the current experiment, which has a single chamber. In the current design the two measurements with electric fields reversed must be made at different times, and therefore any uncorrected magnetic field drifts will show up as noise in the measurement. In the proposed arrangement there are four cells with a common **$B_o$** field in which measurements are made simultaneously. The centre two cells have equal and opposite electric fields, while the outside two cells have no applied electric field. The **+E** and **–E** nEDM measurements are therefore made simultaneously in adjacent cells, and thus any shift in their common **$B_o$** field would induce false nEDMs of opposite sign in the two cells. The two outside zero-field cells act as neutron magnetometers to monitor any changes or gradients in the magnetic field with a level of sensitivity which is (by construction) the correct one for monitoring the nEDM measurement. We are also considering an alternative design which would split the central HV electrode into two with a $5^{th}$ cell in between (which would therefore also have no

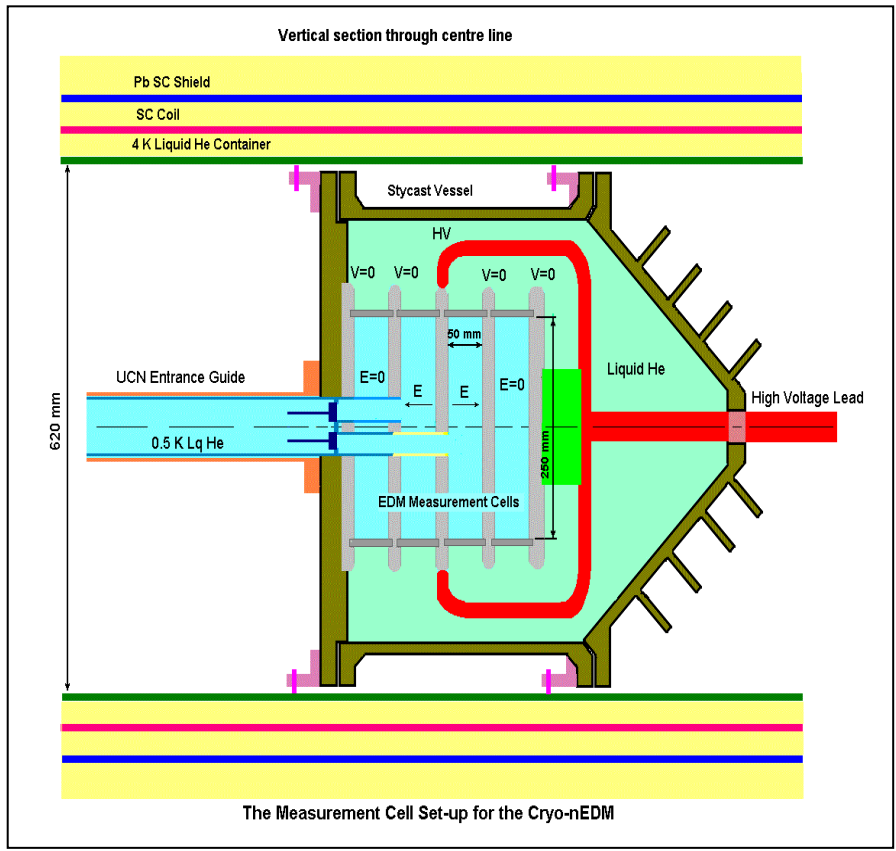

The Measurement Cell Set-up for the Cryo-nEDM

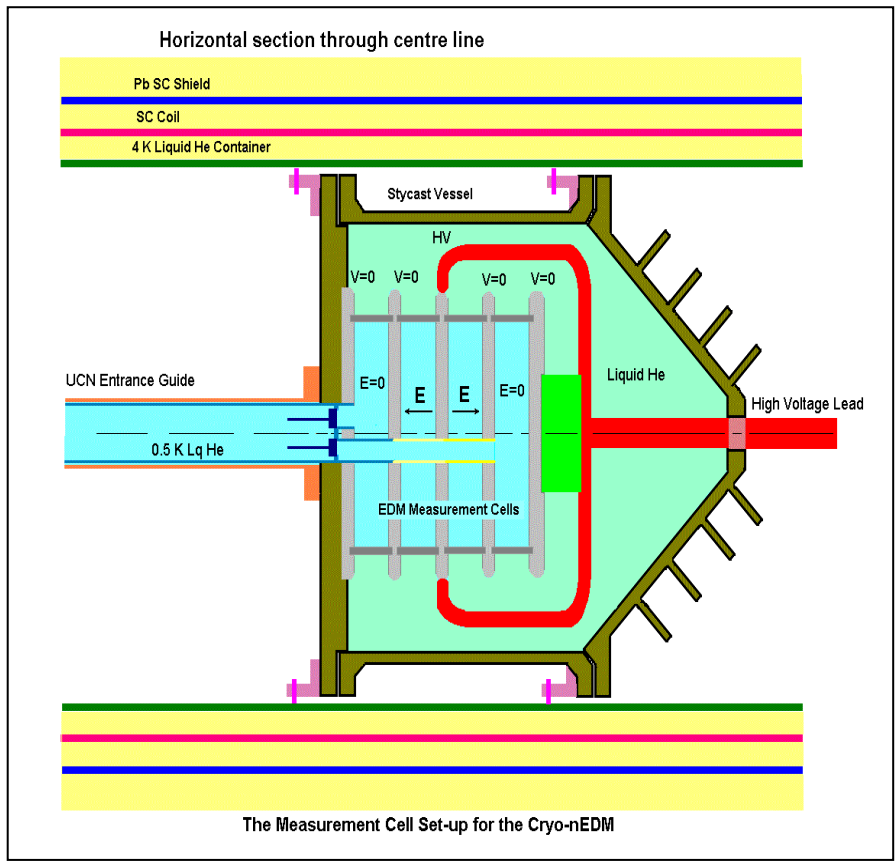

The Measurement Cell Set-up for the Cryo-nEDM

electric field). This would give an even better handle on B-field gradients, at a cost of some additional complexity to the apparatus. The technical complexity of either this 5-cell or the 4-cell system is, however, obviously higher than at present (for instance 4 or 5 valve systems are needed for filling, and the counting cycle becomes rather more complex). This additional complexity is reflected in the system cost as shown in WP5.

## 2.2.5 Controlling and Monitoring the Magnetic Environment

The new cryogenic experiment improves on all four of the parameters that limit the statistical sensitivity of the experiment, but in observing the time-honoured principle that there is no such thing as a free lunch, it does have one drawback. The Hg cohabiting magnetometer, which was so critical to the low level of systematic uncertainties in the room-temperature experiment, cannot be used. We are also leaving behind our tried and testing 4-layer mu metal shield, which did such a good job of reducing the effects of external magnetic fields. It is therefore critical that we have an alternative way of both shielding from external fields and monitoring those fields which do penetrate (or which are produced internally to the experiment). Of course we must also produce the **$B_o$** field in the new apparatus. There are four principal relevant components in the new experiment – the magnetic shielding (mu-metal and superconducting), the superconducting solenoid, the SQUID magnetometry system, and the E-field-free Ramsey neutron magnetometer cells described in section 2.2.4 above.

The shield system (provided by Prof. Yoshiki and the Monbusho) consists of a cylindrical shield of three layers of mu-metal outside a superconducting shield constructed from lead tape. This shield was previously open on both ends, but recently mu-metal endcaps (with a central hole to allow passage of the HV and UCN connections) were added. Calculations imply that this arrangement should provide an adequate shielding factor against external magnetic field disturbances, but this must be verified by testing at Sussex and then *in situ* at PF1 at ILL. Inside this shield is the superconducting solenoid which will be used to produce the uniform magnetic field for the Ramsey measurement.

The total field will be monitored using a SQUID system based on devices designed at Oxford for use on the CRESST experiment. There are a number of differences between this system and the Hg magnetometer now in use. Firstly, of course, a SQUID will work (in fact must work) at cryogenic temperatures. Secondly, the Hg circulates throughout the cell during the period of the measurement and measures the average field over the cell volume (in fact there is a slight difference between the field experienced by the Hg and the neutrons because the neutrons sag about 3mm under gravity that must be allowed for in the analysis). Each individual SQUID channel, however, measures the field integrated over the loop of wire connected to it. In order to measure the field over a volume one must therefore have many channels, and we are in fact proposing to build a 16 channel system in order to get the degree of sampling we believe to be necessary. Another important feature of the new design is the zero-field cells, as these (hopefully) will verify that such fluctuations are absent. A useful number to keep in mind is that if we reach our design UCN density and push the polarisation product to 0.9, the statistical fluctuations in a single filling of a 2.5 litre cell would produce fluctuations in the derived nEDM which are the same size as that produced by a magnetic field fluctuation of 1.0 nG over the 300s measuring period. Therefore if magnetic field fluctuations are small compared to this they will not have any significant effect on the measurement (as long, of course, as they are uncorrelated to the E-field changes), and so random magnetic field fluctuations less than or of the order of 0.5 nG can be ignored. If fluctuations were present at around the 1 nG/300s scale the SQUID magnetometers and the zero-field cells would allow them to be corrected for (with some loss in sensitivity), the redundant methods giving a valuable check on the systematic uncertainties in such a correction. Fluctuations at a level significantly above 1 nG/300s would begin to noticeably degrade the sensitivity of the experiment and would lead us to either eliminate the source of the fluctuations or improve the shielding.

## 2.2.6 UCN detection, DAQ, and analysis.

The final step in the measurement cycle is the detection of the UCN, and then of course the accumulation and analysis of the data. The actual detection is carried out using silicon solid state detectors coated with a thin layer of $^6$LiF. The alphas and tritons from $^6$Li(n,$\alpha$)t then trigger the solid state detector. The RAL group has just completed measurements under a PPARC blue-skies research grant demonstrating that such detectors can be made to work and survive *in situ* in superfluid LHe (C.A. Baker *et al.*, accepted for NIM A (2003). We have recently moved beyond these measurements to demonstrate that such detectors with an added thin film (~1500 Å) of magnetically polarised iron (see Figure V) can be used as spin analysers, as will be required in the new experiment.

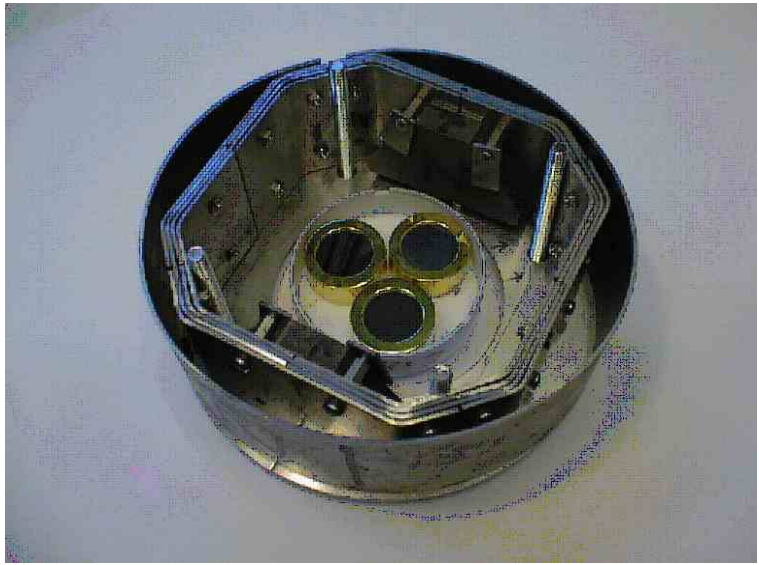

**Fig. 16:** Three 300 mm$^2$ ULTRA detectors used for spin polarisation analysis are set within an array of permanent magnets

Some further development (as covered in Work Package 2) of this detection system is necessary in order to reach the desired sensitivity. Firstly, the detectors have to be made bigger (~35 cm$^2$) so that they will come as close to possible to completely filling the bottom of the UCN detection tube. Without this change the UCN which fall to the bottom of the tube but miss the detector have a high loss probability, which reduces our counting efficiency. Secondly, we need to reduce the noise in our detectors/electronics so that we can clearly see the alpha peak as well as the triton peak, allowing our current detection efficiency of ~41% to be raised to ~90% (where we are normalising the efficiency to a $^3$He detector), an improvement with an effect equivalent to more than doubling the stored UCN density.

In addition to improvements to the detectors themselves it will be necessary to upgrade some of the rather venerable electronics (>20 year old home-brew boxes made from obsolete and unobtainable parts with no remaining spares) and some of the software. This is covered in Work Package 7.

## 2.2.7 Expected Statistical and Systematic Sensitivity

How much better sensitivity to a non-zero nEDM can we expect with this new apparatus? The question divides into two parts: firstly, what is a reasonable estimate of the limiting systematic uncertainty, and secondly, will we have the measurement sensitivity to reach that level? The first part of this question is addressed in detail in Appendix 1, where all the sources of systematic uncertainty which have been included in the design considerations are listed along with detailed numerical estimates of their size. The accuracy of these estimates is controlled by the accuracy of our estimates of the limiting performance of certain physical parameters of the apparatus – leakage currents, displacements, etc. In some sense the table of systematic uncertainties could be viewed as a table of the design goals for those parameters in order to reach the desired sensitivity rather than a table of sensitivities given known values of those parameters. However the estimates included for those parameters stem from a combination of engineering judgements, manufacturers specifications, but mostly from years of experience within the collaboration beating the same sources of systematic uncertainties in previous nEDM measurements, and seem to us to be readily achievable. It should be noted when considering the future of this experiment beyond the period covered by this proposal that none of these parameters is up against any physical limit, and that the bottom line for systematic sensitivity quoted in Appendix 1 is by no means the ultimate limit of the technique. That bottom line number (see Table A1.1) is in fact $1.7 \times 10^{-28}$ $e$ cm, an improvement of about two orders of magnitude over our existing sensitivity.

Will we have the measurement sensitivity to reach this systematic limit? This question is addressed in Appendix 2, which shows the expected sensitivity improvements which will arise from various upgrades to the apparatus. These are rather harder to be precise about, because they depend on quite a number of quantities which we have yet to measure and can only estimate. Certainly we can see a clear path consisting of a number of successive steps each of which would improve the error limit as defined in Equation 2. The question that we cannot answer with certainty at this point is: How many of these steps will be necessary to reach $\sim 10^{-28}$ $e$ cm? Just to take one example, the first item in Table A2.1 gives the expected sensitivity increase available from using the high breakdown potential of LHe to double the HV. We are confident of achieving this factor, however the equipment is actually capable of even higher voltages so perhaps an additional factor will be available. We will not know until we conduct HV tests in LHe. The entries in that table range from simple geometric calculations which one can take as essentially certain (like point f.), to quantities where substantial progress is going to happen but exact quantification is currently difficult (like point e.), to quantities where some R&D is still necessary (like point b.). It should also be pointed out that we have a number of ideas, some quite simple (like putting a 180° Bragg reflector for 8.9 Å neutrons at the back of the cryostat so that the cold neutron beam would traverse the source twice), which could give additional factors. We may therefore not get as large an improvement as we expect from some of the individual factors, but on balance we believe that we have been conservative enough in our estimates of the improvements and have enough other factors in hand that we are quite confident that we can achieve the bottom line goal for this construction phase – at the end of this grant period we will have built a cryogenic experiment which when fed the cold neutron flux currently available at the guide exit from the H53 beamline at ILL will achieve a measurement sensitivity of $10^{-27}\ e$ cm in one reactor week.

## 2.3 The exploitation phase and the further future

Given that we will have built such an experiment, what would we do with it during the exploitation phase? The answer to that question will depend very much on where exactly we get to and what we find during this construction phase. It is worthwhile to look again at Table A2.1, in particular at the last two columns which show the number of reactor days required to reach a measurement error of $10^{-27}\ e$ cm and the number of calendar years (at 150 reactor days/calendar year) required to reach $10^{-28}\ e$ cm after various of the enhancement factors from the new experiment have been included in the calculation. A measurement at the $10^{-27}\ e$ cm level should only take a week of reactor time and therefore would be completed during the period of this grant proposal. However a measurement at the $10^{-28}\ e$ cm level would take almost 5 calendar years, and in that case we would probably push ahead with a proposal to get a higher-flux cold neutron beam at ILL. Two ideas have been discussed to achieve this. The first would be to build a new beam line (called H112) off of a currently unused port in the reactor. Calculations indicate that this would produce an increase in the cold neutron beam flux sufficient to reduce the time needed to do a measurement at the $10^{-28}\ e$ cm level to well under one year. Unfortunately space is at a premium so this beam line would have to be rather long (~100m), which is rather expensive in terms of high-quality neutron guides (the estimated price is ~£1M), however this option is still better (and probably cheaper) than taking 5 years to do the measurement. Another option which has been discussed is to use another currently unused port off of the reactor (called IH1) leading via an inclined guide to an experimental location on an elevated platform very near the side of the reactor. This would produce higher fluxes and would almost certainly be much less expensive, however studies of the real feasibility of this idea have not yet been carried out.

We can therefore offer a number of possible scenarios for the exploitation phase of the experiment:

1. While doing the measurements at the ~$10^{-27}\ e$ cm level signals of a non-zero nEDM are found. This, of course, is thought to be the most likely outcome if supersymmetry is a property of nature. In that scenario the first priority is not the determination of the exact value of the nEDM but demonstrating that the signal is really a signature of nEDM and not some experimental artifact. In that case the exploitation phase would be characterized by running the apparatus in many different configurations while changing experimental parameters rather than on any long single runs. It is likely that in that scenario we would be less interested in a new cold neutron beam line than in a rather different nEDM experiment, perhaps based on vertical extraction of UCN out of our new source into our old room-temperature experiment (which could be upgraded to not be systematics limited at a ~few X $10^{-27}$ *e* cm), to use as a check of the apparatus described in this proposal.

2. The numbers in Table A2.1 turn out to be conservative and we reach a flux higher than the estimate by a factor of 2 or more, but no sign of a non-zero nEDM has been seen at the end of the three-year period covered by this grant. In that scenario we would probably opt to simply continue running the experiment for a couple of years until the measurement sensitivity approached the systematic limit. If no signal had been seen by then we would have to re-evaluate and see if small modifications to our apparatus could reduce the systematic limit and if the data rate was sufficient to make that worthwhile.

3. The numbers in Table A2.1 turn out to be accurate or perhaps slightly optimistic, and no sign of a non-zero nEDM is seen at the end of this three year grant. In that scenario it would probably be most cost-effective to try to increase the cold-neutron flux incident on the experiment in order to get the statistics needed to reach the systematic limit (although the numbers actually quoted in Table A2.1 are probably pretty close to the borderline case – measurement sensitivies better than that and you would just run, worse than that and you would probably opt to upgrade the cold neutron beamline). We would then probably propose one of the ILL upgrades discussed above (while, of course, continuing to run the experiment).

## 3. Schedules and Costs
Deleted for the public version of this document.

## 4. Management Plan
Deleted for the public version of this document.

## 5. Risk Analysis
Deleted for the public version of this document.

# Appendix 1:

# Systematic error analysis for the Cryo-nEDM measurement

Set out in the following few pages are the results of analysing at all the known sources of systematic error in relation to this proposal. The list of ways in which such errors could arise has become fairly stable following 25 years of evolution and the list below has not changed for the last five years. However, the huge increase in the sensitivity of the measurements means that nearly all of the nine mechanisms are sufficiently close to causing trouble that one must consider them carefully and make concessions to them in designing and operating the apparatus.

## Summary comments

As the results in Table A1.1 on the next page suggest, the hardest cases are the second and third and last. The issue of cell displacements under high voltage forces requires very careful design to maintain an appropriately rigid locating of the cells relative to the super-conducting $B_0$-field coil. The problem has been relieved to some extent by approximate cancellation of the results of displacements that come from using a double cell system. The third item – electrical leakage requires careful design and construction to maintain as much symmetry as possible in the leakage current circuit. Again the double cell system gives some useful cancellation. The last effect listed, the geometric phase effect, makes it highly desirable to have the two outer neutron magnetometer cells. This problem is also reduced in proportion to $1/B_0^2$. Indeed, some of the other effects are reduced by increasing $B_0$, notably the second order $\mathbf{E} \times \mathbf{v}$ effect and the effect of 50 Hz ripple on the high voltage. However, the results of 10 kHz ripple will be increased.

All possible steps will be taken as in the past to reduce artefacts and provide diagnostics. For example, the following are alternated in a regular pattern – the side of the Ramsey resonance minimum used for the working point, two positions used on each side of the resonance minimum, the sign of the E-field and sign of the $B_0$-field. Comparison of these different measurements provides a powerful disriminant against sources of a false nEDM. It should also be noted that the numbers in the table all depend upon certain assumptions about the physical parameters of the as-of-yet unbuilt and untested apparatus – leakage currents, displacements, etc. We have used what seem reasonable and in fact conservative estimates of these parameters. These could turn out to be rather better than the estimates, in which case we could reach even lower levels of systematic uncertainty. Conversely there could be unanticipated problems in achieving the goals listed here for those parameters, which would result in redesign leading to schedule slippage.

| Mechanism | False EDM Uncertainty | Assumptions |
|---|---|---|
| Non-zero $(B_0\uparrow\uparrow - B_0\uparrow\downarrow)$ from mu-metal hysteresis | $10^{-2} \times 10^{-28}$ e cm | $(B_0\uparrow\uparrow - B_0\uparrow\downarrow)$ outside the super-conducting shield is that previously experienced in our nEDM experiments |
| Electric forces - cell displacement - $dB_0/dr$ | $1.0 \times 10^{-28}$ e cm | $dB_0/dr = 3\times10^{-8}$ G/mm Rigidity of radial displacement of cells = 100 kg/mm |
| Electrical leakage currents caused by **E** | $1.0 \times 10^{-28}$ e cm | Current of 1 nA at 40 kV/cm An asymmetric tangential flow of 50 mm |
| DC B- and E-fields directly from the high voltage supply | $10^{-5} \times 10^{-28}$ e cm | DC current 1 mA in 40 cm diameter circuit 1.6 m from the shield end – current reverses with sign of HV |
| AC B-fields from the high voltage and dE/dt | $0.05 \times 10^{-28}$ e cm | Ripple on the high voltage 0.04 % - manufacturers figure. 10 kHz and 50 Hz considered. |
| $(\mathbf{E} \times \mathbf{v})/c^2$  1st order UCN ensemble translation of CM | $0.2 \times 10^{-28}$ e cm | Upwards displacement of the UCN due to warming in storage = 1 mm. Volume ave. angle **E** to $\mathbf{B_0}$ = 0.1 radian |
| $(\mathbf{E} \times \mathbf{v})/c^2$  1st order UCN ensemble net circulation about CM | $0.3 \times 10^{-28}$ e cm | **Circulation decay $\tau$ = 1s** $\Delta E_r = E/10$ in outer 30 mm UCN enter at $R/4$ 2s wait before $1^{st}$ $\pi/2$ flip |
| $((\mathbf{E} \times \mathbf{v})/c^2)^2$  2nd order affects all individual trajectories | $0.3 \times 10^{-28}$ e cm | Gives $E^2$ shift $(E\uparrow - E\downarrow)/<E> = 0.05$ $<E>$ = 60 kV/cm used Two cells cancel effect to 10% |
| $(\mathbf{E} \times \mathbf{v})/c^2$ & $dB_0/dz$ geometric phase affects all individl. trajectories | $0.8 \times 10^{-28}$ e cm | $dB_0/dz = 1$ µG/m after trimming. $B_0$ = 25 mG Rms v (UCN) = 5 m/s |
| ***Overall systematic error*** | $1.7 \times 10^{-28}$ e cm | All the above errors are uncorrelated |

Table A1.1    Summary of systematic error estimates with assumptions

# False EDM signals from hysteresis in the mu-metal shield

A broad class of false EDM signals can arise from changes in the $B_0$ field interacting with the neutron magnetic moment where these $B_0$ changes are caused by reversing the E-field. The relevant measurement relation is:

$$h(\nu_{\uparrow\uparrow} - \nu_{\uparrow\downarrow}) = |2\mu_n| (B_{0\uparrow\uparrow} - B_{0\uparrow\downarrow}) - 4d_n E$$

where the arrows represent the directions of the B-field and E-field respectively. (One notes that the neutron magnetic moment $\mu_n$ is a negative quantity). If the first term on the right is significant, then any error in its assessment will be attributed to the second term and thereby give rise to a false EDM. Table X below shows size of the uncontrolled B-field change $(B_{0\uparrow\uparrow} - B_{0\uparrow\downarrow}) = (2d_n E / |\mu_n|)$ that must occur on reversing the E-field in order to create false EDM signals of various sizes.

| E-Field | 10 kV/cm | 20 kV/cm | 40 kV/cm | 60 kV/cm |
|---|---|---|---|---|
| $1\times10^{-26}$ e cm | $3.2\times10^{-11}$ G | $6.4\times10^{-11}$ G | $1.3\times10^{-10}$ G | $1.9\times10^{-10}$ G |
| $1\times10^{-27}$ e cm | $3.2\times10^{-12}$ G | $6.4\times10^{-12}$ G | $1.3\times10^{-11}$ G | $1.9\times10^{-11}$ G |
| $1\times10^{-28}$ e cm | $3.2\times10^{-13}$ G | $6.4\times10^{-13}$ G | $1.3\times10^{-12}$ G | $1.9\times10^{-12}$ G |

Table X   B-field changes correlated with the sign of **E** that give various sizes of false EDM

The nEDM measurement by this group made using separate rubidium magnetometers in 1989 produced false EDMs that were believed to come from hysteresis in the inner layer of the five layer magnetic shield. The mechanism was probably that reversals of **E** induced pickup in the $B_0$ coil circuit. That resulted in transient disturbances of the stabilised current and hence transient changes in the magnitude of **B**$_0$. The $B_0$ flux returns unavoidably through the inner layer of mu-metal. The situation can clearly induce hysteresis in the mu-metal that correlates with the E-field. The false EDMs were of the order $1\times10^{-25}$ e cm when using $E =$ 12 kV/cm. They could only be about half cancelled using the magnetometer readings of $(B_{0\uparrow\uparrow} - B_{0\uparrow\downarrow})$ since these field differences were varying with position and the magnetometers were 15 cm away from the UCN. The nEDM measurement proposed here intends to generate **B**$_0$ using persistent currents in superconductors. This avoids an electronic current stabiliser with its sensitivity to capacitatively coupled voltage pick-up. We will have stabilised trim coils driven in this way, but these will involve less field and will be separated from the measurement cells by a super-conducting magnetic shield tube of length equal to 4 radii. The attenuation of fields in the axial direction propagating into the tube from the ends follows the function $\text{Exp}(-3.83\ z/R)$. For $z/R = 4$ the reduction at the EDM cells is by a factor of $4\times10^6$. For transverse fields the attenuation is $\text{Exp}(-1.85\ z/R)$ and the reduction at the EDM cells is by a factor of $1.5\times10^3$. In this last case field is nominally perpendicular to **B**$_0$. Addition of this small field in quadrature gives only as much affect on $B_0$ as the axial field. But apparatus geometrical asymmetries may put 1 % of the transverse field along **B**$_0$. Then there would be only a $1.5\times10^5$ reduction overall in the change to $B_0$. These attenuations reduce the false EDMs of $1\times10^{-25}$ e cm seen previously with unshielded mu-metal to below $1\times10^{-30}$ e cm, which is negligible. This calculation illustrates one of the big attractions of the super-conducting magnetic shield.

# Displacement of the cell positions in response to electrical forces

We expect small variations in the $B_0$ field strength with position. The gradients in the radial direction $dB_0/dr$ are likely to be of the order $3\times10^{-8}$ G/mm with a correlation length of about 30 cm. It is not easy to monitor or trim absolute gradients $dB_0/dr$. To keep the false EDM signal in one cell below $1\times10^{-28}$ e cm the $B_0$ field change on reversing $E$ must be only $3.3\times10^{-13}$ G. Thus the allowed differential radial displacement of the cell on reversing $E$ is 0.01 microns. Differential movement will only arise from apparatus imperfections. With geometrical symmetry the forces between the high voltage parts and earth parts will have no net radial component. Likewise, for equal strengths of $E$ and $V$, all the forces will be same regardless of the sign of E. We need some idea of the strength of the forces that must balance. The capacitances of the rods etc, are of order 10 pF and their separations are of the order 5 cm. At 50 kV ($E$=10kV/cm) the charges are $5\times10^{-7}$ C and the force between charges of this magnitude at 5 cm separation is 1 newton or 100 gms weight This scale of forces increases as the square of the E and V values used. If there is a 5 % asymmetry in the magnitude of V with sign we are dealing with a differential in the forces of 10 gms weight. If the forces themselves are asymmetric by 10 % due to apparatus imperfections the net radial differential force might be 1 gm weight. We will design the rigidity of the supports to be such that 100 kg weight will produce 1 mm of displacement. On this basis 1 gm weight will produce 0.01 microns of displacement. The effect of the $B_0$ changes, to the extent that they correlate between the two cells, will give opposite signs of false EDM in the two cells. On averaging over the two cells the surviving false EDM is likely to be reduced to $2\times10^{-29}$ e cm. Nevertheless, this will increase in proportion to the E–field that can be used reaching $1.2\times10^{-28}$ e cm at $E$ = 60 kV/cm. This study shows that the correct mounting of the measurement cells is very important. They should be rigidly attached to the base of their surrounding Stycast liquid He containing vessel. The vessel must in turn be located in the super-conducting shield container tube with mountings that have just enough and not too much rigidity, i.e. 100kg/mm of displacement, since too much rigidity could cause damage when there is differential thermal expansion.

# Electric leakage currents driven by the high E-field

The B-field created by the leakage current $i$ may be calculated using the standard result $d\mathbf{B} = (i\,\mathbf{ds} \times \mathbf{r})/(4\pi\,r^3)$ where $\mathbf{r}$ is the vector from the current element $i\,\mathbf{ds}$ and the position for $d\mathbf{B}$. In first order, we only need to obtain the contributions $dB_z$ where $z$ is the symmetry axis of the stack of storage cells along which lies the main field $\mathbf{B_0}$. Only current elements in a tangential direction about the cylinder axis $z$ can contribute to $dB_z$. Of course the E-field is also nominally aligned with $z$ but imperfections in the cell storage insulating wall and/or the placement of the electrode connections may cause the current to have a net displacement in a tangential direction. We will suppose that this displacement is 50 mm and that there is a cancellation caused by spill over of the $dB_z$ into the adjacent cell where it gives the opposite sign of false EDM. We estimate a reduction factor of 5 from this averaging over the two cells. Then the leakage currents, which give the fields in the above table, are:

| E-Field kV/cm | 10 kV/cm | 20 kV/cm | 40 kV/cm | 60 kV/cm |
|---|---|---|---|---|
| $1\times10^{-26}$ e cm | 25 nA | 50 nA | 100 nA | 150 nA |
| $1\times10^{-27}$ e cm | 2.5 nA | 5.0 nA | 10 nA | 15 nA |
| $1\times10^{-28}$ e cm | 0.25 nA | 0.5 nA | 1.0 nA | 1.5 nA |

Typical leakage currents in the existing room temperature EDM experiment are 1 nA. Such currents are expected to be much less in super-fluid helium due to the absence of 'channeltron' type multiplication of field emission that is characteristic of insulators in E-fields in vacuum. We do however anticipate a new phenomenon - an additional current of the order of 0.1 nA caused by ionisation generated in the helium of the cell by the half MeV electrons from decay of the stored UCN (end point energy 720 keV).

## DC B- & E-fields emanating directly from the high voltage supply

For its feedback and stabilisation this kind of supply generally uses a bleed current of about 1 mA through a high resistance as a measure of the high voltage achieved at the top of the Cockroft-Walton stack. Taking a diameter of 40 cm for this circuit we estimate the field components (axial and transverse) at the end of the main magnetic shield 1.6 m away to be $2\times10^{-8}$ G. In propagating further to influence $B_0$ at the measurement cells we have the following attenuations for the transverse component - feedback coils $10^{-1}$, mu-metal shield $10^{-2}$, SC shield $10^{-5}$, partial cancellation between cells $2.5 \cdot 10^{-1}$, which after doubling for a reversal of the current with reversal of HV sign gives a B-field change of $1\times10^{-16}$ G and an EDM signal, when working at 40 kV/cm, of $1\times10^{-33}$ e cm. The axial component produces less effect still due to a 100-fold greater attenuation in the SC shield.

The greatest danger from voltages in the high voltage supply is contamination of other parts of the equipment through common earths, etc. As we have done previously we will arrange that the high voltage supply is, DC-wise, as isolated as possible from all other circuits. In the previous experiment the HV supply had its very own computer and all exterior communication with the supply was via fibre optic cables. The 300 kV line itself is of course an exception, but we will control the current flow during the Larmor precession hopefully by using a connecting cable that is a giant photodiode – a semiconductor at 0.5 K, where the conduction is controlled by light supplied through a fibre optic within the cable.

## Effects of oscillatory B-fields linked to the E-field

The high voltage source will have ripple at some level that will give rise to an alternating displacement current through the cells. This will generate in the cells an AC B-field in the xy-plane that will cause a shift in the Larmor frequency. How this shift is calculated depends on the ripple frequency. The drive frequency of the Cockroft-Walton stacks is 10 kHz – well above the Larmor frequency of 75 Hz. In this case, there is a downward shift in the Larmor frequency equivalent to a B-field change given by $\Delta B_0 = -B_{ac}^2 B_0/(2B_{fac}^2)$ Here $B_{ac}$ is the strength of the AC B-field and $B_{fac}$ is the frequency of the AC field multiplied by the field to Larmor frequency conversion factor $(2\pi/\gamma)=(1G/3.0$ kHz$)$. If we would like to keep our false EDM below $1\times10^{-28}$ e cm at E = 10 kV/cm or less than $1.5\times10^{-29}$ e cm at 60 kV/cm then we must keep the magnitude of $2\Delta B_0$ down to $3.2\times10^{-13}$ G. At 10 kHz, $B_{fac}$ = 3.3 G and we plan to use $B_0 = 2.5\times10^{-2}$ G allowing $B_{ac}$ to be $1.7\times10^{-5}$ G. If this is the field 13 cm from a line current, then the current must be $1.1\times10^{-3}$ A. The cell has a capacity of 10 pF and an impedance, at 10 kHz, of $8.5\times10^5$ ohms, so the 10 kHz voltage giving this current is 1.3 kV. At 60 kV/cm the total voltage will be 300 kV so the ripple can be 0.6 %. In fact the specification of the supply available is 0.04 %.

We now take the case of ripple at 50 Hz. This, being a little below the Larmor frequency adds approximately in quadrature to the main field and increases the Larmor frequency $\Delta B_0 = B_{ac}^2/(2B_0)$. For the same criterion on false EDMs, $B_{ac}$ is in this case $1.3\times10^{-7}$ G and the allowed AC current $8.4\times10^{-6}$ A. The 10 pF cell at 50 Hz has an impedance of $3.2\times10^8$ ohms so the voltage allowed is 260 V. At 300 kV, 260 V is 0.08 % and again this is within the specification of the available HV supply. Thus we can tolerate a difference in ripple of this amount between positive and negative high voltage. Finally, any false EDM from this effect reverses sign when the $\mathbf{B_0}$ field is reversed in the apparatus because the particular high voltage polarity that has the higher ripple will create parallel **E** and **B₀** fields for one direction of **B₀** and antiparallel **E** and **B₀** fields for the opposite direction of **B₀**. Similarly there is cancellation between the back-to-back measurement cells, in that case because they have opposite directions of **E**, but the same direction of **B₀**, for a given high voltage polarity. These cancellations should give us at least a factor of ten reduction to $1 \times 10^{-28}$ e cm.

The high voltage supply will also generate AC B-fields directly and these will typically be three orders of magnitude stronger than its DC B-fields. The AC field reaching the cells could be $10^{-2}$ G$\times(2 \times 10^{-3})(10^{-1})(10^{-2})(7 \times 10^{-4}) = 1.4 \times 10^{-11}$ G, However AC B-fields can only affect $B_0$ by adding in quadrature giving $2\Delta B_0 = 1 \times 10^{-21}$ G. This is negligible being about eight orders of magnitude less than we need concern ourselves with.

# First order E×v effect

Motion through the E-field causes an UCN to see the derived field $\Delta \mathbf{B} = (\mathbf{E} \times \mathbf{v})/(c^2)$ in tesla. If there exists a finite component $\Delta B_z$ this will reverse with reversal of **E** and introduce a false EDM signal corresponding to the resulting $2\Delta B_z$ and the results in the above Table X. We are concerned with the ensemble average over the batch of UCN and the interval of time over which the Larmor frequency measurement is made. The measurement starts when the first π/2 flip of the Ramsey sequence is made and finishes after a time laps of $T$ with the second π/2 flip. In calculating the magnitude of **v** we recognise that the the motion of any group of particles (here the batch of UCN) can be separated into the motion of the centre of mass and the motion of the members relative to the centre of mass. Accordingly, the 1ˢᵗ order (**E×v**) effect has two parts:

## *Translation of the centre of mass*

This is dominant in Beam experiments but is much suppressed in trap experiments. The relevant velocity $\mathbf{v} = (\mathbf{r}_1 - \mathbf{r}_2)/T$, where $\mathbf{r}_1$ and $\mathbf{r}_2$ are the positions of the centre of mass of the batch of UCN at the beginning and end of the Ramsey interval $T$. Differences in position can arise as follows:

(a) The UCN may warm slightly in storage due to Doppler shifts from vibrating walls or by scattering of nanoparticles moving with Brownian motion on the wall surface. Due to gravity the UCN centre of mass starts off displaced downwards by of the order of 10 mm from the geometric centre of the cell. The amount of depression is inversely proportional to the UCN energy. Warming has been observed at the level of 10 % of their energy after a long interval of storage and this would reduce the depression of the CM by 1 mm. The resulting velocity is 1 mm/300 s = $3 \times 10^{-6}$ m/s. It is vertically upwards nominally perpendicular to the horizontal **E**. If **E** is parallel to **B₀** the derived field $\Delta \mathbf{B}$ will be perpendicular to **B₀** and $\Delta B_z$ = zero. The issue then is what small angle is likely to exist between **E** and **B₀**. We expect our B-field to be under control to 1 part in 2000. The problem is more likely to be caused by the E-field where the angle between the volume averages of **E** and **B₀** might be as much as 0.1 radian due to the charging of the insulator. This leads to $2\Delta B_z = 7 \times 10^{-14}$ G at $E$ = 10 kV/cm with $\Delta B_z$ being proportional to $E$. This gives a false EDM of $2 \times 10^{-29}$ e cm independent of E and is quite safe.

(b) Just at the end of filling with UCN there can be an initial transient off-set of the CM as a memory of the filling process. The issue is the time taken for the UCN to become distributed uniformly over the cell. The time for UCN to cross the cell is about 0.04 s. The filling time constant is likely to be about 5 s, so most UCN will have crossed the cell about a 100 times by the time the door is closed. Of course, they are near the door when they enter. Some simple analysis in which they are assumed to all enter simultaneously shows that for smooth and physically reasonable distributions of $\upsilon_x$ between zero and the $\upsilon_{x\,max}$ the expectation for the CM position is at less than $W_x/(8n+1)$ from the central equilibrium position after UCN with $\upsilon_{x\,max}$ have $n$ transits of the width $W_x$. If $W_x$ is 100 mm, a value $n = 13$ ensures a displacement of less than 1 mm. If $\upsilon_{x\,max}$ is 6 m/s this takes less than ¼ second to achieve and we conclude that a 1 s delay before the first π/2 flip will ensure that the false EDM from this cause is even less than the $2\times10^{-29}$ e cm of part (a) above.

### *Net circulation of UCN in one particular sense about the centre of the cell*

The batch of neutrons may acquire some net circulation (angular momentum) about the centre of the cell in the filling process. At the same time there could be a finite $E_r$ everywhere outwards in the outer 30 mm of the radius of the cell through charging of the insulator. Net circulation of the UCN through this peripheral region will result in a $\Delta B_z$ from **E**×**v** as long as the circulation continues. The walls will have a matt finish and the net circulation should decay in a few seconds due to irregular reflection of the UCN.

If the UCN enter off-centre in the cell they can acquire angular momentum from the force of gravity in the interval before their centre of mass has moved to the centre of the cell. The displacement from the centre is a decaying oscillation with a decay constant of about 0.04 seconds and a period of about 0.08 seconds. The tangential velocity acquired at the periphery is about ($g$. 0.04 s)/4 = 0.1 m/s. (The ¼ is an estimate of the direction cosine into the tangential direction at the periphery for particles starting R/4 from the centre – a direction cosine that goes to zero if the UCN start exactly at the centre of the cell. The decay constant for loosing this angular momentum is the time for about 10 collisions with the matt walls i.e, about 1 s. Thus averaged over the 300 s measurement interval the velocity is 0.1 m/s ×(1 s / 130 s) = $7.7\times10^{-4}$ m/s. If the $E_r$ field change on reversing E is E/10 we find $2\Delta B_z = 1.7\times10^{-11}$ G at E = 10 kV/cm and proportional to E. However, this distortion of E is only present in that part of the cell volume within 30 mm of the side wall, or about 1/2 of the volume, taking us to a volume average $2\Delta B_z$ of $8\times10^{-12}$ G. This corresponds to a false EDM of $2.7\times10^{-27}$ e cm, which is intolerable. However, we need to consider the decay of the angular momentum in the delays before the measurement begins. If the filling of the cell continues for 2.3 times the filling time time constant (2.3×5 s = 11.5 s) only about 5 % of the UCN will enter in the last 2.5 s and those already in the cell will have lost about 95 % of their angular momentum so we have another factor of 10 reduction to giving a false EDM of $2.7\times10^{-28}$ e cm. It is easy to add a few seconds of delay following cell door closure before applying the first π/2 flip pulse to further reduce the angular momentum of the UCN and that is perhaps the most important conclusion of this analysis.

## Second order E×v effect

Of course the main strength of the derived field $\mathbf{B_v} = (\mathbf{E}\times\mathbf{v})/c^2$ given that **E** and $\mathbf{B_0}$ are nominally aligned, is in the xy-plane perpendicular to $\mathbf{B_0}$. Given a constant $\mathbf{B_v}$ this addition increases the strength of the overall $B_0$-field (from Pythagorous) by the amount $B_v^2/(2B_0) = E^2\upsilon_{xy}^2/(2B_0c^4)$ and increases the Larmor frequency proportionally. Of course the UCN keep bouncing on the walls. After each bounce, $\mathbf{B_v}$ adopts a new direction in the xy-plane but keeps a similar magnitude. Provided the bounce frequency is much less than the Larmor precession frequency this time structure has little effect for this calculation and the above formula remains a good approximation for the small increase in the $B_0$-field strength with its associated increase in the Larmor frequency. The shift is seen to be proportional to $E^2$ and nominally does not give any false EDM signal. In practice reversal of the E-field may be accompanied by some small change in its magnitude, e.g, 5 % in $E$ and 10 % in $E^2$. Then the equivalent B-field shift between the two directions is given by $(2\Delta E/E)(E^2 v_{xy}^2)/(2B_0 c^4) = 5\times 10^{-13}$ G at 10 kV/cm. From Table X, this translates into a false EDM of $1.6\times 10^{-28}$ e cm at 10 kV/cm and $1\times 10^{-27}$ e cm at 60 kV/cm. However, there is further protection with the back-to-back cell system driven by the same power supply because the higher strength E-field is for **E** anti-parallel to **B$_0$** in one cell and **E** parallel to **B** in the other cell. Finally, any false EDM from this effect reverses sign when the **B$_0$** field is reversed in the apparatus because the particular high voltage polarity that has the higher E will, in any one cell, create parallel **E** and **B$_0$** fields for one direction of **B$_0$** and antiparallel **E** and **B$_0$** fields for the opposite direction of **B$_0$**. This cancellation is expected to be at least as good as 10 % giving an average false EDM 40 times smaller than the above figures e.g. $4\times 10^{-30}$ e cm at 10 kV/cm and $3\times 10^{-29}$ e cm at 60 kV/cm.

## Geometric phase effect

There are two kinds of small field components appearing in the xy-plane. There are those caused by a gradient in the $B_0$-field, which for cylindrical symmetry take the form **B$_r$** = (d$B_0$/dz)(**r**/2). And there are the fields **B$_v$** generated by the **E×v** transformation. If we call these **a** and **b** respectively then they act together in the form (**a**(t) + **b**(t)). All the Larmor frequency shifts produced these $B_{xy}$-fields, even when time structure of the fields matters, depend on the square of the total xy-field acting, so we are concerned with terms **a**$^2$, 2**a.b** and **b**$^2$. The **b**$^2$ term is just the 2$^{nd}$ order **E×v** effect of the previous section. This section concerns the cross term 2**a.b**. The term **a**$^2$ does not involve the E-field and is not of interest unless we worry about an indirectly brought about change in **a$^2$** as a response to reversing **E**. This would be a small effect in a small effect. Any such influences are much more likely to be seen through changes in the main component $B_{0z}$.

For the geometric phase case 2**a.b**, the two types of field collaborate to form a field that the moving UCN see as rotating in a definite sense in the xy-plane even though the particles move isotropically in the cell. We are preparing a publication on the false EDM signals caused by geometric phases for particles in traps. Here we will just quote an analytic result that has been confirmed by numerical simulations. When the principal $B_0$-field distortion is described by a finite d$B_0$/dz and the UCN collision frequency is much less than the Larmor frequency the false EDM in e cm units is = (100/6) (ℏ/e) ((d$B_0$/dz)/$B_0^2$)(< $v^2$>/$c^2$) where SI units must be used. In combating this effect, it is important to make d$B_0$/d$z$ smaller than its initial value – i.e, that value of about $3\times 10^{-5}$ G/m which is expected in the apparatus without any $B_0$-field trimming. When the spectrometer is operated with UCN, the Larmor precession results from the adjacent measurement cells will give a measurement of the gradient present. The gradient, monitored thus, can be reduced using trim coils. A 30-fold reduction to $1\times 10^{-6}$ G/m can easily be obtained in this way. The false EDM given by the above formula is then $8\times 10^{-28}$ e cm. This is a slightly idealised approach. With less assumptions about the smoothness of the $B_0$-field, one can show that one should use, in place of d$B_0$/d$z$, the number of flux lines which are emerging from the side walls of the two cells that have entered through the up-field electrode. This number of lines must be divided by the area of the side walls times the radius of the cell. The number of lines can be obtained from the difference in average fields seen by the two neutron magnetometer cells built into the electrode plates. Thus one would both trim the $B_0$-field as suggested and monitor the situation using the two magnetometer cells. There is no cancellation of this false effect between the two EDM measuring cells. One notes that the false EDM is inversely proportional to $B_0^2$ so it helps to keep $B_0$ stronger if other things allow it.

**False EDM signals from artefacts**
The analyses above are answering the question how can the high voltage supply and/or the E-field influence the <u>actual</u> neutron Larmor precession rate. In this section we consider how the high voltage could cause a change in other apparatus parameters that are used to calculate the Larmor precession rate from the UCN counts at the end of the batch cycle and thereby influence the frequency result via the calculation. This includes changes in the counts themselves that might be caused by pick-up from the high voltage in the UCN detector channel. The naïve way to carry out the experiment is to pick a working point on the steepest point to one side of the central minimum of the Ramsey resonance curve for the spin-up counts curve of Fig. Y page xx. The gradient of the steepest part of the centrl loop of the curve, where the point counts of the curve are for single batch cycles, is $\pm \pi\alpha<N>T$ UCN counts/Hz where the minus sign is for the low frequency side of the minimum and the plus sign for the high frequency side. For the present argument we choose just the latter. An EDM will manifest itself by giving a lower spin-up count when the **E** and **B$_0$** fields are parallel and a higher count when they are antiparallel. The frequency difference between the two field directions for an EDM of $1\times10^{-28}$ e cm at E = 40 kV/cm is $4\times10^{-9}$ Hz. The gradient with $\alpha=0.7$, $<N> = 1.4\times10^5$ and T=300s is close to $1\times10^8$ counts/Hz so the difference between the spin up counts for the two E-field directions for this EDM is 0.4 counts for the pair of batch cycles. Thus a difference of 1 count per UCN batch cycle coming from noise on specifically the positive high voltage setting would give a false EDM of $2.5\times10^{-28}$ e cm. Small count differences could also from tiny shifts in the discriminator level. The high voltage supply will be kept as separate as possible from the rest of the system, but it is difficult to give it an entirely independent mains supply. Another possibility is some link between the high voltage and the cycle timing control. Since the surviving neutrons decreasing steadily throughout the batch cycle, a timing change will change the count - one part in $10^5$ change yielding 1.4 counts. Our approach is to preload, at the start of a one day run, a microprocessor controlled quartz clock timer that then runs autonomously throughout the run. There may still be small effects from valve action once the command to shut a valve is launched, but since this is only a matter of 3 in $10^4$ of the cycle time making tiny alterations in the valve response time to be negligible. In any case for this new experiment the valves will be moved by step motors, which give little scope for small analogue influences.

Our principal protection from the class of biases just outlined is to elaborate on the naïve procedure by operating the batches alternately between the two sides of the central resonance minimum exploiting the fact that the slopes there are equal in magnitude and opposite in sign. The only thing needed to make this change of side is to load the frequency synthesiser driving the flip coils of the Ramsey system with a slightly different number for the frequency requested. We can think of no reason why this digital change leading to a frequency change of 1 part in 1000 in an oscillating field of only $10^{-4}$ G should have the slightest effect on the count difference between the two batches with different directions of **E**. The different signs of slope cause the count difference to translate into different signs of false EDM so that they then cancel. One could now ask just how equal in magnitude are the two slopes? The Ramsey resonance shape is a $\sin^2$ function of frequency with a half period of 1.7 mHz with a bell shaped overall envelope function with a full half-width of 400 mHz. The only shift of the envelope pattern from the centre of the fringes is that casued by the Block-Seigert effect because we use a linear oscillating, rather than a rotating, flip field. It is about 0.1 mHz. Without this offset, the slopes of the (unit amplitude) envelope function at $\pm$ 0.85 mHz from the centre are, assuming the central part is parabolic in shape, equal to $\pm (0.85/400)^2 = \pm 4.5\times10^{-6}$. The envelope offset changes these to about $+5.5\times10^{-6}$ and $-3.5\times10^{-6}$ Hence the symmetry of the central loop of the resonance curve is broken at the level of only $2\times10^{-6}$. Thus, if a spurious ten thousand counts difference per pair of batch cycles with opposite directions of **E** was present (7 % of the expected true counts) correlated with the sign of the E-field, the resulting false EDM after averaging over the two sides of the resonance curve would still only be $5\times10^{-30}$ e cm. Spurious counts on this scale would be easily detectable by quick and simple test measurements on the apparatus. It is our conclusion and our past experience that false EDM signals due to artefacts are relatively easy to defeat.

# Appendix 2: Measurement Sensitivity

The nEDM error due to neutron counting statistics noise alone is:

$$\sigma(d_n) = \frac{\hbar}{2\alpha ET\sqrt{N}}$$

where in the current room temperature experiment, using a $Hg^{199}$ co-magnetometer in a 20.8 litre EDM storage cell, on the PF2 UCN source at ILL:

α = polarization/analysis product
T = neutron storage time in the measurement cell
E = electric field
N = number of neutrons detected

**Room Temperature nEDM**

In the current room temperature experiment, using a $Hg^{199}$ co-magnetometer in a 20.8 litre edm storage cell, on the PF2 UCN source at ILL:

α = 0.64 with lifetime $\tau_{pol}$ ~ 450 s
T = 130 s with lifetime $\tau_{ucn}$ ~150 s
E = 10 kV.cm$^{-1}$
N = 14,000 UCN/edm cycle of 215s

With these parameters the measurement sensitivity is:

$$\sigma(d_n) = 17 \times 10^{-26} \text{ e.cm /day}$$

**Cryogenic nEDM**

With the same polarization product; electric field and UCN stored lifetime a cryo-edm experiment at the present H53(PF1) position at ILL and a 'L shaped' experimental arrangement would give:

$$\sigma(d_n) \sim 21 \times 10^{-26} \text{ e.cm /day}$$

since

1. UCN production in superfluid $He^4$ = (0.35 ± 0.05) UCN/cm$^3$/s with $10^7$ neutrons/cm$^2$/s/Å at 8.9 Å
2. $\Phi_c$ (capture flux at 8.9Å) = 3.0 x $10^7$ n.cm$^{-2}$.s$^{-1}$ with reflecting spin filter/polariser in beam
3. UCN detection efficiency = 41% (only tritons observed)
4. Spin Polarisation UCN ~ 100% x beam polarisation

and assuming  a) UCN lifetime in source region $\tau_{ucn}$ = 300s
b) UCN transfer efficiency = 20 % (with source/transfer/edm cell = 5/14/5 litres)

**Table A2.1 Parameter improvements and resulting statistical sensitivity**:

|  | Current value | Reasonably expect |  | Gain in Sensitivity | edm/day x $10^{-26}$ e.cm | Reactor days to reach $1 \times 10^{-27}$ e.cm | Calendar years to reach $1 \times 10^{-28}$ e.cm |
|---|---|---|---|---|---|---|---|
| **EDM cell** | | | | | | | |
| Electric field (E) | 10 kV.cm$^{-1}$ | 20 kV.cm$^{-1}$ | a | 2.0x | 10.5x$10^{-26}$ | 11,025 | 7,350 |
| Polarization product ($\alpha$) | 0.6 | 0.9 | b | 1.5x | 7.0 | 4,900 | 3,267 |
| Storage time (T) | 130 s | 300 s | a | 1.8x | 3.9 | 1,521 | 1,014 |
| | | | | | | | |
| **Neutron factors** | | | | | | | |
| UCN detection efficiency | 41% | 90% | c | 1.5x | 2.6 | 676 | 451 |
| H53 beam flux at 9A | $\Phi$=2.6 x $10^7$ n.cm$^{-2}$.s$^{-1}$.A$^{-1}$ | $\Phi$=1.0 x $10^8$ n.cm$^{-2}$.s$^{-1}$.A$^{-1}$ | d | 2.0x | 1.3 | 169 | 113 |
| transmission polariser | 20% | 50% | e | 1.5x | 0.9 | 81 | 54 |
| Beam/He$^4$ areas | 24% | 100% | f | 2.0x | 0.45 | 20 | 13 |
| UCN density dilution source to edm cell | 20% | 50% | g | 1.5x | 0.3 | 9 | 6 |
| narrow/broad band beam | 75% | 100% | h | 1.1x | 0.26 | 7 | 4.7 |
| | | | | | | | |
| **New Beamline** | | | | | | | |
| H53/H112 neutron beam | 15% | 100% | i | 2.6x | 0.10 | 1 | 0.7 |

N.B. 150 reactor days = 1 calendar year

**Enabling factors**:

a). properties of liquid He$^4$
b). rebuild spin analyzer with higher B field + R&D on spin retaining materials
c). detect both $\alpha$ and tritons from  n+ Li$^6$ => $\alpha$ +t
d). put neutron guide between H53 exit and experiment
e). build new polariser using sapphire substrate optimized for 9A neutrons
f). match beam area to area of UCN containment source
g). either increase UCN source volume or R&D on low volume (small tubes) UCN transfer system
h). remove velocity selector but this will increase the activation of the apparatus
i). New guide at ILL can be built but will cost £1.0M (Farhi & Malbert's design/estimates).